# On-chip Integration of High-Frequency Electron Paramagnetic Resonance Spectroscopy and Hall-Effect Magnetometry

H. M. Quddusi, C. M. Ramsey, J. C. Gonzalez-Pons, J. J. Henderson and E. del Barco

*Physics Department, University of Central Florida, 4000 Central Florida Boulevard, Orlando, Florida 32816-2385*

G. de Loubens and A. D. Kent

*Department of Physics, New York University, 4 Washington Place, New York, New York 10003*

**ABSTRACT**

A sensor that integrates high sensitivity micro-Hall effect magnetometry and high-frequency electron paramagnetic resonance spectroscopy capabilities on a single semiconductor chip is presented. The Hall-effect magnetometer was fabricated from a two-dimensional electron gas GaAs/AlGaAs heterostructure in the form of a cross, with a 50×50 $\mu m^2$ sensing area. A high-frequency microstrip resonator is coupled with two small gaps to a transmission line with a 50 $\Omega$ impedance. Different resonator lengths are used to obtain quasi-TEM fundamental resonant modes in the frequency range 10-30 GHz. The resonator is positioned on top of the active area of the Hall-effect magnetometer, where the magnetic field of the fundamental mode is largest, thus optimizing the conversion of microwave power into magnetic field at the sample position. The two gaps coupling the resonator and transmission lines are engineered differently—the gap to the microwave source is designed to optimize the loaded quality factor of the resonator ($Q \leq 150$) while the gap for the transmitted signal is larger. This latter gap minimizes losses and prevents distortion of the resonance while enabling measurement of the transmitted signal. The large filling factor of the resonator permits sensitivities comparable to that of high-quality factor resonant cavities. The integrated sensor enables measurement of the magnetization response of micron scale samples upon application of microwave fields. In particular, the combined measurement of the magnetization change and the microwave power under cw microwave irradiation of single-crystal of molecular magnets is used to determine of the energy relaxation time of the molecular spin states. In addition, real time measurements of the magnetization dynamics upon application of fast microwave pulses are demonstrated.

## I. INTRODUCTION

The control and manipulation of spin and charge degrees of freedom of nanoscale systems [1-3] has become a major challenge during the last few years in part due to potential applications in quantum computation and quantum information technologies. In particular, molecular nanomagnets [4-13] have become prototype systems to probe the borders between quantum and classical physics, as well as to study decoherence in quantum systems [14,15]. However, in most experimental studies the quantum dynamics of the magnetic moment of the molecules is governed by incoherent quantum tunneling (i.e., the characteristic time of the experiment is longer than the decoherence time). Only recently, have experimental groups [16-21] attempted to study magnetization dynamics upon application of electromagnetic radiation in the microwave range. Although these experiments provided very exciting results and have become the milestones in the study of relaxation phenomena in molecular magnets, they lacked the experimental conditions required to perform time-resolved studies of decoherence of molecular magnets in their solid state form. On the technical side, one of these requirements is the ability to generate large ac magnetic fields at the sample under study (i.e. efficient conversion of the microwave power into ac magnetic field at the sample position). Therefore, the development of novel techniques of measurement to identify the relaxation mechanisms in molecular magnets is important toward understanding decoherence in nanoscale magnetic systems and to explore the potential of molecular magnets in quantum technologies.

In this article we present a high-frequency microstrip resonator integrated with a high-sensitivity Hall-effect magnetometer (HEM) that enables the application of large microwave magnetic fields to small samples while providing high-sensitivity detection of the sample response. The sensor allows real-time measurements of the magnetization of micron scale samples in response to microwave excitation, while the microwave power absorbed by the sample can be obtained simultaneously through electron paramagnetic resonance (EPR) spectroscopy. The high efficiency in the conversion of microwave power into ac magnetic field at the sample position together with the fast response of the HEM provide new capabilities for time-resolved studies of quantum dynamics of molecular nanomagnets.

The manuscript is organized as follows: In Section II we present the design and characterization of the microstrip resonators. In Section III, we describe its integration with a two-dimensional electron gas (2-DEG) Hall-effect magnetometer and demonstrate simultaneous measurement of the magnetization and the EPR response of a micron scale single-crystal of $Ni_4$ SMMs upon application of microwave radiation. In Section IV we explore the potential of our integrated sensors for the study of the quantum dynamics and decoherence phenomena in nanoscale magnetic systems. We conclude with Section V in which we summarize our work.

## II. ASSYMETRIC MICROSTRIP RESONATORS

Half-wavelength microstrip line resonators have been used for EPR spectroscopy since 1974 [22,23]. Fig. 1a shows a sketch of a typical reflection microstrip resonator. The fundamental mode ($L \sim \lambda/2$) may be between 5-30 GHz. In our devices, the geometry (i.e., width of the line, $w$, and thickness of the dielectric substrate, $h$, separating the line from the metallic bottom plate) has been calculated to match the impedance of the microwave coaxial lines (50 Ω), to avoid reflections at the device connections.. Note that in our case, where $h < \lambda/2$, the microwave signal transmission corresponds to a quasi-TEM mode.

As seen in the sketch in Fig. 1a, the central line of the reflection resonator is isolated from the microstrip feed line by a coupling gap, $g_c$, which is designed to critically couple the resonator to the feed line, resulting in a loaded quality factor, $Q_L = Q_0/2$, on the order of 100 when critically coupled. Fig. 2 shows the simulated [24] frequency dependence of the reflection parameter $S_{11}$ for reflection resonators with different coupling gaps. The optimal response (critical coupling) is obtained for $g_c = 40$ μm in a resonator with a fundamental mode at ~10 GHz, with the parameters shown in the inset of Fig. 2. High quality factors are required to generate large microwave magnetic fields at the sample location. The dashed gray line in Figure 3a shows the calculated reflection parameter, $S_{11}$, of a reflection resonator with a fundamental resonance frequency of 14.74 GHz. The ac current density at resonance is shown in Fig. 3a (marked as "reflection"), being maximum at the center of the resonator. Correspondingly, the microwave magnetic field is maximized at the center of the resonator.

From saturation measurements of the standard EPR marker DPPH ($S = 1/2$), we have estimated that a power of ~10 mW at the input port of a 10 GHz microstrip resonator ($Q_L = Q_0/2 = 85$) produces an ac magnetic field of $h_{ac}$ ~1 Oe. The field right above the center of the resonator can be written in terms of the input power, $P_{in}$, as follows,

$$h = \frac{\mu_0}{w}\sqrt{\frac{P_{in}Q_0}{2Z}} \quad (1)$$

where $Z$ is the impedance of the resonator ($Z = R$ at resonance). An estimation of the ac field using Eqn. (1) for this resonator gives $h = 1.75$ Oe, in good agreement with the value extracted from saturation. Note that large ac fields are necessary to induce fast Rabi oscillations to overcome decoherence. For a spin $S = 1/2$, an ac-field of 10 Oe, easily attainable in our resonators, would generate Rabi oscillations with a period below 100 ns, a requirement to perform spin-echo experiments.

Reflection resonators are not convenient to work with at low temperatures because long coaxial lines, broken at different places for thermal anchoring, are required to transfer the microwave radiation to the sample. Reflections and standing waves generate oscillations of the reflected power ($S_{11}$) masking the response of the resonator. To solve this problem, we have developed transmission resonators by including a second feed line on the other side of the resonator, separated by a transmission gap, $g_t$, as shown in Fig. 1b. The second feed line may perturb the response of the resonator, increasing the losses and decreasing the quality factor substantially. The black lines in Figure 3a show the reflection $S_{11}$ (solid) and the transmission $S_{21}$ (dashed) parameters of a transmission resonator designed to work at ~ 15 GHz with a transmission gap, $g_t = g_c = 140$ μm. Note that almost 30% of the microwave amplitude (~ 9 % of the power) is reflected back ($S_{11} = 0.3$ at resonance). In addition, the resonance width is substantially larger (i.e. lower Q) than that of the corresponding reflection resonator.

In order to minimize this distortion of the resonator response, we have increased the transmission gap. Fig. 3a shows the $S_{ij}$-parameters of two other resonators with different transmission gap sizes (red and blue lines). Both the transmission and the reflection at resonance decrease (Q increases) upon increasing the transmission gap. Beyond a certain gap size the reflected signal is indistinguishable from that of a reflection resonator ($g_t = \infty$), as is the ac current density (see Fig. 3a), while the transmission is sufficient for measurements. The response of a transmission resonator fabricated on a GaAs substrate with 300 nm-thick gold lines was measured with an Agilent PNA vector network analyzer (see Fig. 3b). We note that our devices are designed for 0.6 mm-thick GaAs wafer.

As stated before, these devices fulfill the technical requirements necessary for the study of quantum dynamics of micron scale solid state magnetic samples. First, the magnetic field is strong and very homogeneous near the center of the resonator (10% variation within 50 µm from the surface), allowing for a very efficient conversion of the microwave power into ac field at the sample position. Second, the input impedance is designed to maximize the power arriving at the resonator. In addition, moderate quality factors (Q ~ 100) provide a fast dissipation of the energy stored in the resonator, allowing the use of fast microwave pulses (> 1 ns) without cavity ringing. Finally, microstrip resonators offer very high sensitivity for susceptibility studies, comparable or even larger than high quality factor resonant cavities (with Q ~ $10^4$). This is due to the fact that the sensitivity is the product of the quality factor and the filling factor, $\eta$. The latter is the ratio of the microwave energy in the sample volume over the total energy stored in the cavity. In microstrip resonators, almost all the energy is concentrated in the proximity of the resonator, at the center of the line, and the sample is within a good fraction of this volume, leading to filling factors on the order of 0.1 (as opposed to $\eta \sim 10^{-3}$ in resonant cavities).

## III. INTEGRATED SENSORS

Another advantage of working with microstrip resonators is that they allow us to integrate EPR spectroscopy with high-sensitivity Hall-effect magnetometry [25]. In Fig. 4a we show a schematic of an integrated sensor. Fig. 4b shows a photograph of an integrated EPR-HEM chip mounted in a low temperature housing box. The chip is made out of a 0.6 mm-thick GaAs/AlGaAs substrate containing a high-mobility 2-DEG layer ~70 nm underneath the surface. The carrier density, $n = 5.46 \times 10^{11}$ cm$^{-2}$, and mobility, $\mu = 67400$ cm$^2$

V$^{-1}$ s$^{-1}$, of our 2-DEG were determined from Hall measurements carried out at $T$ = 77 K.

A 50×50 μm$^2$ cross-shaped HEM has been patterned on the surface by means of optical lithography and 100 nm-deep chemical etch to produce the mesa structure. A zoom at the center of the sensing area of the device is shown in Fig. 4c, where the current and the voltage leads for the operation of the HEM are indicated. The leads open go to the sides of the wafer where macroscopic Au/Ge pads are used to contact micro-coaxial lines for high-frequency (<1 GHz) magnetic measurements at low temperatures (see Fig. 4b). Thermal annealing of the contacts in inert atmosphere was performed to diffuse the germanium into the wafer and obtain ohmic contacts to the 2-DEG layer. A preliminary magnetic characterization of the HEM once the microstrip was patterned on top was carried out at low temperatures (4 K) by applying a magnetic field perpendicular to the sensor plane, while monitoring the Hall voltage across the sensor with a lock-in amplifier in the presence of a low-frequency (< 100 kHz) ac current of 1 μA. The Hall coefficient, $R_H$ = 2,200 Ω/T, was extracted from the dependence of the Hall voltage on the external field and the geometry of the sensor. The density of carriers, n = $1/eR_H$ = 4.2×10$^{11}$ cm$^{-2}$, extracted from the Hall coefficient, agrees well with the value found at 77 K.

The microstrip resonator (described in section II) is patterned on the surface of the device, with the center of the resonator line located on top of the HEM cross, making the area in which the microwave magnetic field of the fundamental resonant mode is maximum coincide with the active area of the HEM. A thin dielectric separates the sensor from the heterostructure material. The ends of the reflection and transmission feed lines are connected to the inner conductor of a 0.086" semi-rigid coaxial line ($f$ < 50 GHz) through opposite walls of the housing box (see Fig. 4b).

To demonstrate the feasibility of our integrated sensor we carried out simultaneous measurements of the magnetization and the EPR absorption spectra of a single-crystal of $Ni_4$ SMMs. For this, a pyramidal shaped crystal (~ 60×60×120 μm$^3$) is placed on top of both sensors as shown in Figure 4c. $Ni_4$ is known to have the easy axes in the direction of the long axis of the pyramid [16]. In our case, a large field is applied transverse to the easy axis to generate tunnel splittings of the order of the microwave radiation energy employed. Thus the crystal is aligned with its easy axis in the direction of the ac magnetic field ($h \perp H_{dc}$). The results are shown in Fig. 5. The magnetization of the $Ni_4$ single crystal recorded at 200 mK during excitation with microwave radiation of $f$ = 15GHz is shown in Fig. 5a (red curve). Deviations from the equilibrium magnetization in the absence of radiation (black curve) are observed at both polarities of the magnetic field, corresponding to photon induced transitions between quantum superposition states of the spin of the molecules (see Ref. [16] for details). The microwave-induced magnetization change is shown in Fig. 5c. Correspondingly, Fig. 5b shows the microwave power absorbed by the sample (estimated from the change in the transmitted power through the resonator, $\Delta S_{21}$, which is recorded simultaneously).

The results presented in Fig. 5 demonstrate that our EPR-HEM integrated sensors allow the simultaneous measurement of the magnetization and the EPR spectra. This is of crucial importance for the study of the magnetization dynamics in nanoscale systems, since the simultaneous measurement of $\Delta M$ and $P_{abs}$, allows the direct determination of the energy relaxation time of the sample, associated with the coupling of the spin and the lattice (i.e. spin-phonon relaxation time). The relation (derived in Ref. [20]) is given:

$$\tau = \frac{hfN_0}{2M_{eq}} \frac{\Delta M}{P_{abs}} \qquad (2)$$

where $N_0$ is the population difference between the two spin levels at equilibrium and is a function of the temperature, $T$, and the total number of molecules in the sample $N_T$. Taking the results shown in Fig. 5, we obtain an energy relaxation time for this sample of ~30ms. In this sample, phonon-bottleneck is known to slow down the energy relaxation of the molecular spin states. This makes the extracted relaxation time larger than the spin-phonon relaxation time of this sample ($\tau > T_1$) [26].

## IV. REAL-TIME MAGNETIZATION DYNAMICS

Our EPR-HEM sensor can also be used to perform time resolved measurements. Using a fast PIN diode switch operating up to 40 GHz and controlled by a pulse generator, it is possible to generate very short (the typical switching time of our pin-switch or pulse generator is 5 ns) microwave pulses. An RF amplifier (with 40 dB gain and 30 dBm maximum output power) amplifies the microwave pulse power after the switch, before it is sent to the microstrip resonator, through the coaxial lines. Using the PNA, it is also possible to detect the microwave power absorbed during such short high power pulses. To these standard pulse EPR capabilities, our integrated sensor adds the capability of time-resolved measurements of the magnetization of the sample during and after microwave pulses. This direct measurement is very useful to investigate relaxation processes, for instance in SMMs [18,19].

Fig. 6 shows an example of such an experiment on a $Ni_4$ single crystal. Here, the microstrip resonator patterned on top of the HEM has a resonance frequency of 26 GHz, and a transverse field of 2.75 T is applied in order to generate the desired tunnel splitting between the two lowest-lying spin-states of the $S = 4$ SMM, as in the experiment presented in Section III. The presented data were obtained at 0.4 K. In order to perform these real-time measurements, a constant DC current of 10 µA is used to bias the HEM, and a low noise preamplifier with a 1 MHz bandwidth is used to amplify the Hall voltage, which is read and averaged on an oscilloscope. The typical Hall voltage is a few µV, and a gain of 1000 is used in order to amplify it to mV levels. The presented data are averaged on a few thousands traces, the W = 2 ms long microwave pulse being there repeated every 100 ms (hence, the duty cycle is DC = 2%). It is possible to record the magnetization behavior during the pulse and after it has been turned off, for different applied longitudinal fields (Fig. 6a). The magnetization decreases during the pulse, and then relaxes towards equilibrium on a few ms time scale. The diminution of the magnetization at the end of the pulse is maximum when the longitudinal field $H_z$ = 0.12 T, *i.e.* when the splitting between the two lowest-lying spin-states matches the microwave energy *hf*. It is also possible to analyze "time slices" of the magnetization curve, as shown in Fig. 6b. In that case, the magnetization changes induced by the microwave irradiation are plotted as a function of $H_z$, at different times during and after the application of the pulse. During the first stage of the pulse (filled circles and squares), the magnetization change is maximum at $H_z$ = 0.12 T, and small for large longitudinal fields, where the resonant condition is not fulfilled. However, large magnetization changes develop even far from resonance, at later times. This is due to direct heating by the microwaves. As the pulse is turned off, the full magnetization curve relaxes towards equilibrium, which is almost recovered after 20 ms, see open triangles. This behavior is well reproduced by a phenomenological model taking into account the heat transfers between the $Ni_4$ crystal, the chip of the integrated sensor, and the bath, as well as a resonant part corresponding to the microwave absorption by the spins around $H_z$ = 0.12 T [27].

Finally we point out that the data presented in Fig. 6 demonstrate a time resolution of about 10 µs, limited in this case by the bandwidth of the amplifier, and that the intrinsic dynamics of the sample studied is much longer, in the ms time scale. Measurements of the magnetization with pulse lengths down to ~ 50 ns nanoseconds have been successfully obtained with our resonator (not shown here). In fact, the performance of the EPR-HEM for time-domain studies is only limited by the wiring and the ohmic contacts to the 2D electron gas, as well as by the electronics employed to amplify the Hall voltage to a level measurable for an oscilloscope.

## V. CONCLUSIONS

We have presented a novel sensor which integrates high-frequency EPR spectroscopy and high-sensitivity Hall-effect magnetometry on a single semiconductor chip. The ability to simultaneously monitor the behavior of the magnetization and the absorbed microwave power of a sample provides an efficient tool to study the dynamics of nanoscale magnetic systems. Furthermore, the integrated sensors have great potential for novel studies of the quantum dynamics of molecular nanomagnets by allowing real-time monitoring of the magnetization evolution upon application of fast microwave excitation pulses.

**ACKNOWLEDMENTS**

E.d.B. acknowledges support from the US National Science Foundation (DMR0706183 and DMR0747587) and technical assistance from Agilent Technologies. G.d.L and A.D.K acknowledge support from the US National Science Foundation (DMR0506946, DMR0703639) and the Army Research Office (ARO-MURI).

**FIGURES**

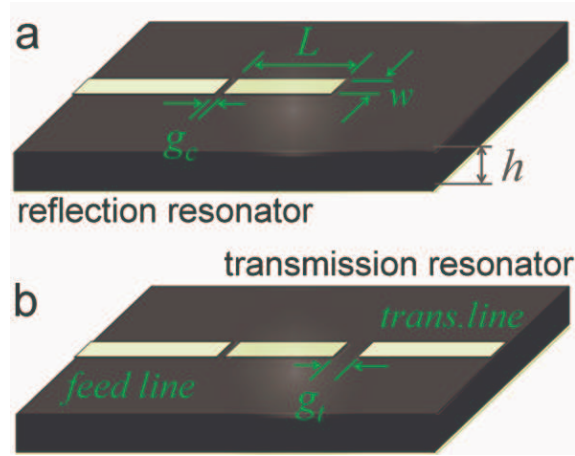

**FIG.1:** (Color online) a) Reflection and b) transmission microstrip line resonators. The coupling and transmission gaps are engineered differently to optimize device performance.

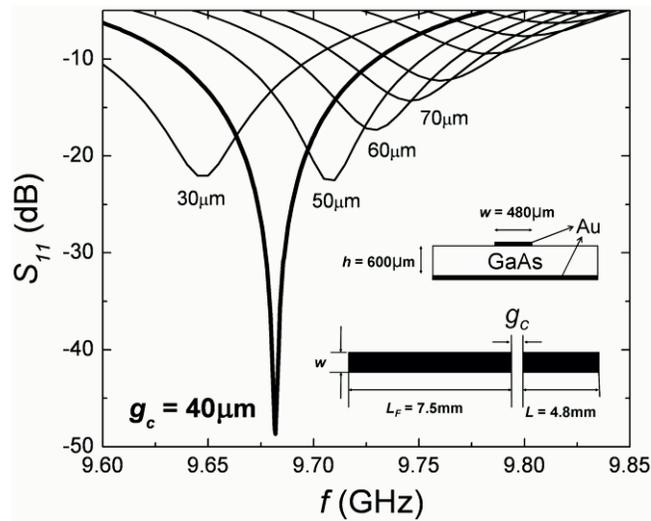

**FIG. 2:** $S_{11}$ as a function of frequency for different coupling gaps between the feed line and the central line of a ~10 GHz microstrip resonator.

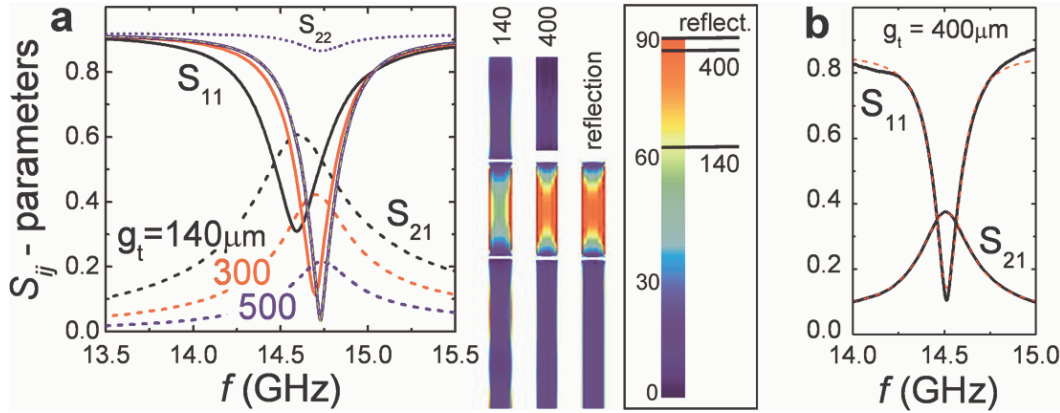

**FIG. 3:** (Color online) a) Reflection and transmission parameters of ~15 GHz transmission resonators with several transmission gaps. The color-coded sketches on the right represent the magnitude of the ac current in the device at resonance. b) Measured response of a resonator fabricated on GaAs.

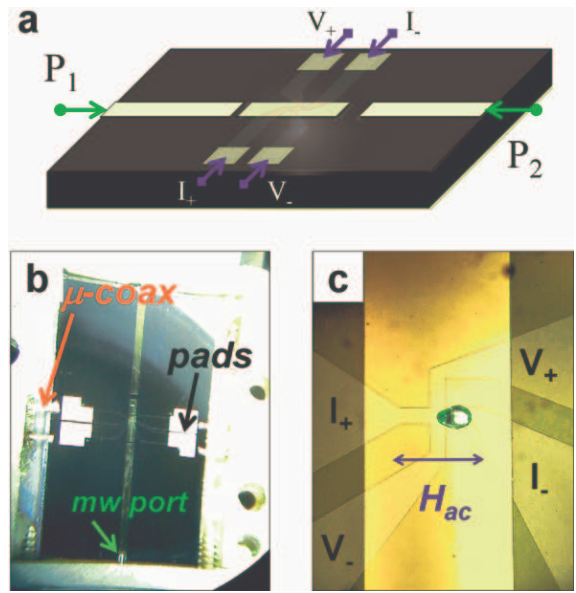

**FIG. 4:** (Color online) a) Sketch of an EPR/HEM integrated sensor. b) Photograph of the device placed in a low temperature housing box. c) Optical micrograph of the center part of the resonator. The 50×50 $\mu m^2$ cross-shaped HEM can be seen underneath the resonator. The green pyramidal-shaped sample is a single crystal of a $Ni_4$ SMM.

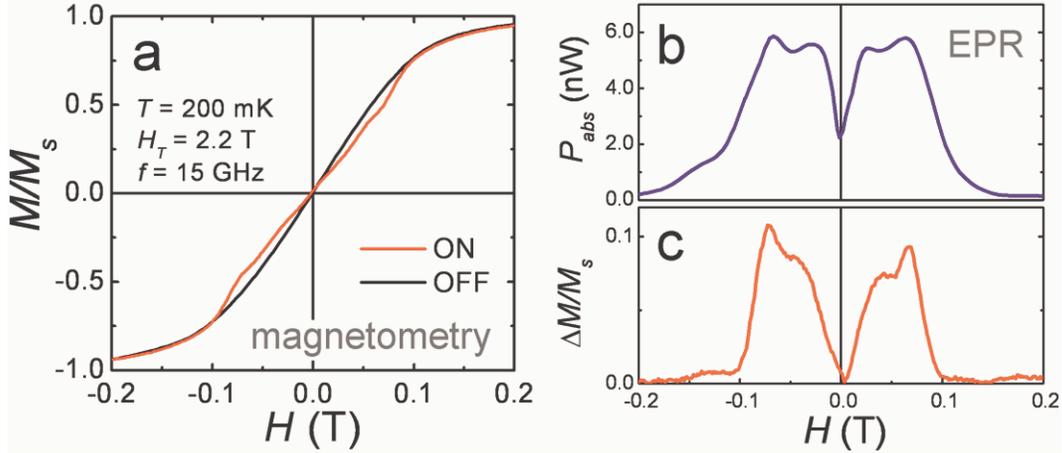

**FIG. 5:** (Color online) a) *Black curve*: Magnetization response of the Ni$_4$ SMM crystal at 0.2K as a function of easy axes magnetic field (a constant transverse field of 2.2T is also applied). *Red curve*: The magnetization under 15 GHz microwave excitation applied to the sample. b) Microwave power absorbed, extracted from the signal transmitted through the resonator, S$_{21}$. c) Microwave-induced change of magnetization extracted from (a).

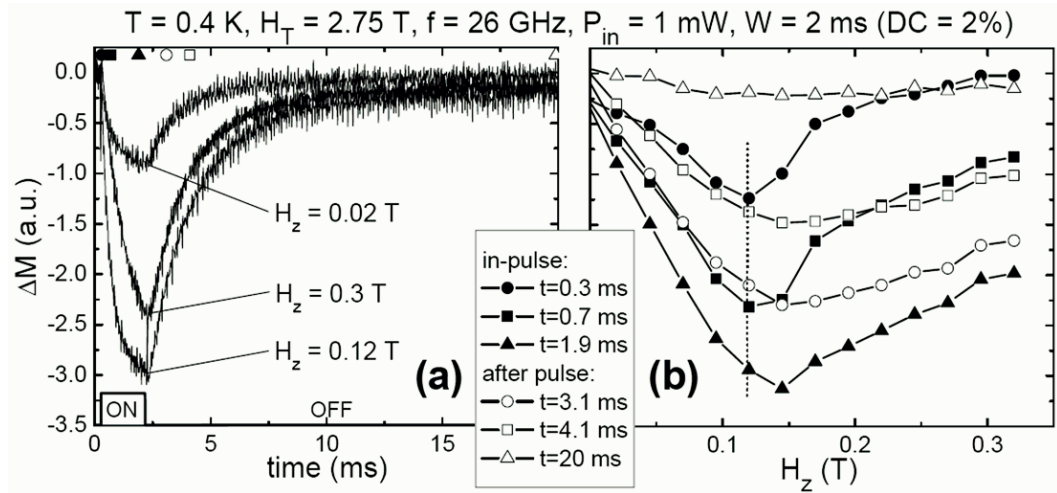

**FIG. 6:** (a) Time-resolved magnetization measurements for 3 different applied longitudinal fields under microwave pulse irradiation. (b) Magnetization curves at 6 different times, during and after the microwave pulse. The dashed line indicates the resonant field corresponding to the transition between the 2 lowest-lying tunnel split spin-states.

17